\documentclass[10pt,journal,compsoc]{IEEEtran}
\usepackage{amsmath}
\usepackage{amssymb}
\usepackage{graphicx}
\usepackage{subcaption}
\usepackage{amsmath}
\usepackage{amsthm}
\usepackage{multirow}
\usepackage{booktabs}
\usepackage{algorithm}
\usepackage{algorithmic}
\usepackage{bm}
\usepackage{enumitem}
\usepackage{xspace}

\newcommand{\ie}{\emph{i.e.,}\xspace}
\newcommand{\eg}{\emph{e.g.,}\xspace}
\newcommand{\et}{\emph{et al.}\xspace}

\usepackage{amssymb}
\def\B#1{\mathbf #1}
\def\C#1{\mathcal #1}
\def\R#1{\mathbb #1}

\ifCLASSOPTIONcompsoc
  \usepackage[nocompress]{cite}
\else
  \usepackage{cite}
\fi
\ifCLASSINFOpdf
\else
\fi
\begin{document}
%

\title{Pre-training Graph Transformer with Multimodal Side Information for Recommendation}

%
%

\author{
    Yong~Liu,
    Susen~Yang,
    Chenyi~Lei,
    Guoxin~Wang,
    Haihong~Tang,\\
    Juyong~Zhang,
    Aixin~Sun,
    and~Chunyan~Miao

    \IEEEcompsocitemizethanks{
        \IEEEcompsocthanksitem Yong Liu is currently with Alibaba-NTU Singapore Joint Research Institute and Joint NTU-UBC Research Centre of Excellence in Active Living for the Elderly (LILY), Nanyang Technological University, Singapore 639798. Email: stephenliu@ntu.edu.sg.
        \IEEEcompsocthanksitem Susen Yang and Juyong Zhang are currently with School of Mathematical Sciences, University of Science and Technology of China, Hefei, Anhui, China 230052. Email: susen@mail.ustc.edu.cn, juyong@ustc.edu.cn.
        \IEEEcompsocthanksitem Chenyi Lei is currently with Department of Electronic Engineering and Information Science, University of Science and Technology of China, Hefei, Anhui, China 230027, and the Alibaba Group, HangZhou, Zhejiang, China. Email: chenyi.lcy@alibaba-inc.com.
        \IEEEcompsocthanksitem Guoxin Wang is currently with College of Biomedical Engineering \& Instrument Science, Zhejiang University, Hangzhou, Zhejiang, China 310007, and the Alibaba Group, HangZhou, Zhejiang, China. Email: xiaogong.wgx@taobao.com.
        \IEEEcompsocthanksitem Haihong Tang are currently with the Alibaba Group, Hang Zhou, China. Email: piaoxue@taobao.com.
        \IEEEcompsocthanksitem Aixin Sun and Chunyan Miao are currently with School of Computer Science and Engineering, Nanyang Technological University, Singapore 639798. Email: axsun@ntu.edu.sg, ascymiao@ntu.edu.sg.

    }

    \thanks{Manuscript received xxx, 2021; revised xxx, 2021.}
}

\IEEEtitleabstractindextext{%
\begin{abstract}
Side information of items, \eg images and text description,  has shown to be effective in contributing to accurate  recommendations. Inspired by the recent success of pre-training models on natural language and images, we propose a pre-training strategy to learn item representations by considering both item side information and their relationships. We relate items by common user activities \eg co-purchase, and construct a homogeneous item graph. This graph provides a unified view of item relations and their associated side information in multimodality. We develop a novel sampling algorithm named MCNSampling to select contextual neighbors for each item. The proposed Pre-trained Multimodal Graph Transformer (PMGT) learns item representations with two objectives: 1) graph structure reconstruction, and 2) masked node feature reconstruction. Experimental results on real datasets demonstrate that the proposed PMGT model effectively exploits the multimodality side information to achieve better accuracies in downstream tasks including item recommendation, item classification, and click-through ratio  prediction. We also report a case study of testing the proposed PMGT model in an online setting with 600 thousand users.


\end{abstract}

\begin{IEEEkeywords}
Recommendation Systems, Graph Neural Networks, Self-Supervised Learning, Pre-Training model
\end{IEEEkeywords}}

\maketitle

\IEEEdisplaynontitleabstractindextext

%
\IEEEpeerreviewmaketitle

\IEEEraisesectionheading{\section{Introduction}\label{sec:introduction}}

\begin{figure}
\centering
\includegraphics[width=0.98\columnwidth]{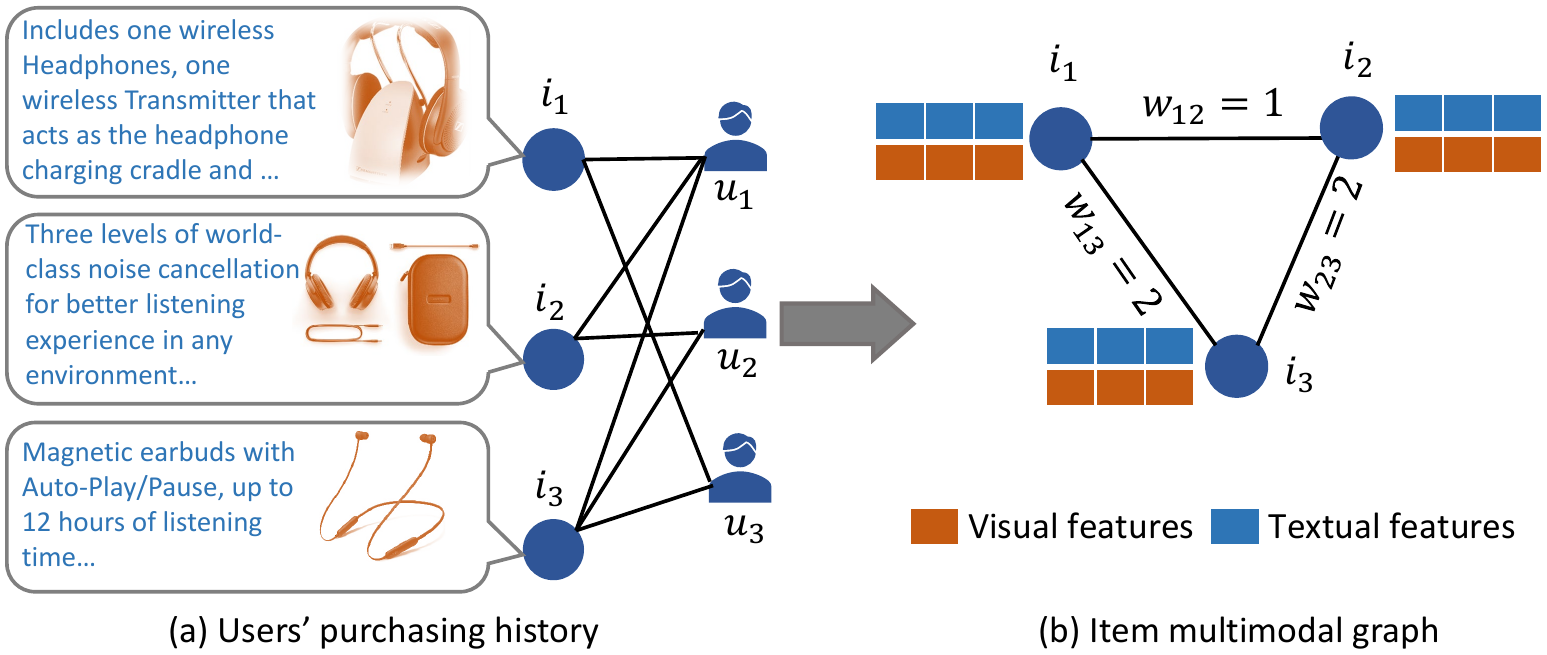}
\caption{(a) Item side information and users’ purchasing history, and (b) Item multimodal graph built on co-purchase relationship. In this graph, each node denotes an item with its visual and textual features extracted from the image and text description respectively. An edge between two items is weighed by the number of co-purchases.}
\label{fig:itemgraph}
\end{figure}

\IEEEPARstart{P}{ersonalized} recommendation systems have attracted significant attentions in recent years from both industry and academic. A good range of techniques have been proposed, from the classic collaborative filtering techniques, \eg matrix factorization~\cite{shi2014collaborative}, to the recent deep learning models~\cite{zhang2019deep}, \eg convolutional neural networks~\cite{tang2018personalized} and recurrent neural networks~\cite{hidasi2015session}. 
In addition to the interactions between users and items, the multimodality side information of items  has also been exploited and showed effectiveness in  further improving recommendation accuracy~\cite{sun2019research}. Example side information of items include textual descriptions, images, and videos, as shown in Figure~\ref{fig:itemgraph}(a).


Traditional methods exploit item side information by manual feature engineering~\cite{liu2016repeat}, and then employ factorization machine~\cite{rendle2010factorization} or gradient boosting machine~\cite{chen2016xgboost} to predict users' preferences on items. These methods often require domain-specific knowledge, and are time-consuming. Deep learning-based methods leverage the strong representation learning ability of neural networks to exploit item side information, for learning the user and/or item representations.
However, existing solutions only consider a specific type of side information of items for the dedicated recommendation applications~\cite{he2016ups,chen2018neural,wu2019context}. The full multimodality side information of items are not fully exploited.


\begin{figure*}
\centering
\includegraphics[width=1\textwidth]{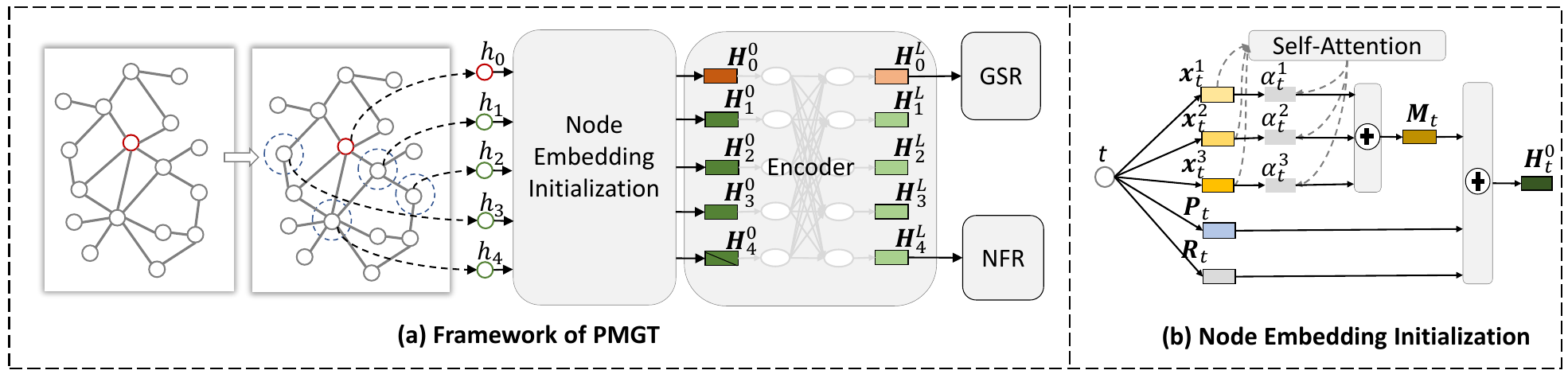}
\caption{(a) An overview of the proposed PMGT framework. PMGT contains four components (illustrated from the left to the right): contextual neighbors sampling, node embedding initialization, transformer-based graph encoder, and graph reconstruction. GSR and NFR, in the last step, denote the graph structure reconstruction task and the masked node feature reconstruction task, respectively. (b) The node embedding is initialized by considering the node's multimodal features, position-id embedding, and role-label embedding.}
\label{fig:framework}
\end{figure*}

Inspired by the successes of unsupervised pre-training strategies designed for natural language processing~\cite{devlin2018bert} and graph data~\cite{hu2019strategies,hu2020gpt,zhang2020graph}, we propose to develop an unsupervised pre-training framework to fully exploit the multimodality side information for item representations learning. Illustrated in Figure~\ref{fig:itemgraph}, we construct an \textit{item multimodal graph} to provide a unified view of items with their associated  multimodality side information. In this item multimodal graph (see Figure~\ref{fig:itemgraph} (b)),  each node is an item and the edges model their relationships (\eg co-purchase or co-viewership, depending on the application domain).  We then pre-train graph neural network (GNN) on this item multimodal graph to enable the GNN model to capture both item relationships and their multimodality side information.

Note that, relationship in the item multimodal graph is based on user interactions with items, which is independent of content similarity between items. We do not manually define content similarity by items' textual and/or visual features because our aim is to learn item representations partially on top of these multimodal features. Moreover,  item relationships by user interactions also play an important role in recommendation. For example, ItemKNN~\cite{sarwar2001item} is a representative collaborative filtering approach that exploits the item similarity based on their common user interactions. It has been shown to be even more effective than a few neural recommendation methods in some experiments~\cite{dacrema2019we}. For matrix factorization methods, the regularization term defined based on the item similarity by user interactions also help improve recommendation accuracy~\cite{han2020conextual}.


The unified item multimodal graph well distinguishes our work from  previous studies where  side information of items and their relationships are studied separately. We argue that these two types of information complement each other in solving recommendation problems. We hence focus on effective pre-training on top of this item multimodal graph to benefit item recommendation. We also show that our pre-training  benefit other E-commerce applications like click-through ratio (CTR) prediction.

The contributions made in this paper are as follows. First, we propose a novel pre-training framework, namely Pre-trained Multimodal Graph Transformer (PMGT), to exploit items' multimodality information through unsupervised learning. To the best of our knowledge, this is the first deep pre-training method developed to exploit the multimodality side information of items for recommender systems. Second, we decompose the learning objective of PMGT into two sub-objectives: (i) graph structure reconstruction, and (ii) masked node feature reconstruction. To handle large-scale graph data, we develop an algorithm, named Mini-batch Contextual Neighbors Sampling (\ie MCNSampling), for effective and scalable training. Moreover, we employ the attention mechanism to aggregate the item's multimodality information, and a diversity-promoting Transformer framework to model the influences between an item and its contextual neighbors in the graph. Lastly,  to demonstrate the effectiveness of the proposed PMGT model, we conduct extensive experiments on real datasets for different applications. The results on three downstream tasks, \ie item recommendation, CTR prediction, and item classification, demonstrate that PMGT is more effective than existing graph-based pre-training methods in fully exploiting items' multimodality information. To further show the effectiveness of PMGT, we report a case study of applying this model in an online E-commerce platform.



The rest of this paper is organized as follows. Section~\ref{sec:relatedwork} reviews existing pre-training strategies in different domains. Section~\ref{sec:model}  details the proposed PMGT framework. Next, Sections~\ref{sec:recexperiments} and~\ref{sec:expitem} present the experimental results of three downstream tasks. Then, in Section~\ref{sec:case}, we show a case study of applying the proposed model in an online E-commerce platform. Lastly,  Section~\ref{sec:conclusion} concludes this paper.

\section{Related Work}
\label{sec:relatedwork}

Our work is related to pre-training strategies developed for computer vision (CV) and natural language processing (NLP) tasks, and also  pre-training methods for graph structure data. We briefly review the related studies in these two areas.

\subsection{Pre-training Methods for Vision and Language}

Pre-training methods have been widely applied in CV and NLP tasks~\cite{he2019rethinking,devlin2018bert}. It has been shown that pre-training is effective in boosting  performances of various downstream applications. For example, a general pre-training paradigm for CV tasks is firstly training a model on the ImageNet dataset~\cite{deng2009imagenet}, and then fine-tuning the pre-trained model for a specific  task. Self-supervised learning~\cite{jing2020self} has recently been applied to pre-train vision models~\cite{goyal2019scaling}. NLP is another domain where pre-training is usually adopted to learn word representations. The shallow pre-training methods, \eg word2vec~\cite{mikolov2013distributed} and GloVe~\cite{pennington2014glove}, learn word representations based on the word co-occurrence patterns in a corpus of documents. Recently, significant progress has been made in developing deep pre-training models for contextual word representations. For example, ELMo~\cite{peters2018deep} employs a bidirectional language model to learn high-quality deep context-dependent word representations. The BERT~\cite{devlin2018bert} and XLNET~\cite{yang2019xlnet} models use attention mechanisms to learn the word representations. Significant improvements are achieved when applying these pre-trained models on various NLP tasks.

\subsection{Pre-training Methods for Graph}
On graph data, many embedding techniques have been developed in recent years, and we refer the readers to~\cite{cai2018comprehensive} for a comprehensive survey. Representative shallow graph embedding methods include TransE~\cite{antoine2013trane}, DeepWalk~\cite{perozzi2014deepwalk}, LINE~\cite{tang2015line}, and Node2vec~\cite{grover2016node2vec}. The recent popularity of GNNs motivates the development of pre-training strategies for GNN models. In general, these methods pre-train GNNs by solving the graph reconstruction problem. For example, Kipf \et introduce the variational graph autoencoder (VGAE) framework for graph reconstruction~\cite{kipf2016variational}. Hamilton \et propose a general inductive framework called GraphSAGE~\cite{hamilton2017inductive}, which exploits node features to generate node embeddings by sampling and aggregating features from a node's local neighborhood. Velickovic \et propose the Deep Graph Infomax model~\cite{velickovic2019deep} that aims to maximize the mutual information between the node representations and the representation of the graph. Recently, self-supervised learning~\cite{jing2020self} is employed to simultaneously pre-train GNNs at both node and graph levels, and an example is the self-supervised learning method for graph neural networks~\cite{hu2019strategies}. Similarly, the generative framework GPT-GNN~\cite{hu2020gpt} employs a self-supervised attributed graph generation task to pre-train GNNs, by effectively capturing both semantic and structural properties of the graph. In~\cite{qiu2020gcc}, a self-supervised graph neural network pre-training model is proposed to capture the universal network topological properties across multiple networks. In~\cite{zhang2020graph}, the proposed Graph-BERT model employs attention mechanism to aggregate the neighborhood information of a target node in the graph.


\section{The Proposed Pre-training Model: PMGT}
\label{sec:model}

Figure~\ref{fig:framework} shows the proposed PMGT framework. Observe that PMGT contains four main components (illustrated from the left to the right in Figure~\ref{fig:framework} (a)): 1) contextual neighbors sampling, 2) node embedding initialization, 3) transformer-based encoder, and 4) graph reconstruction. Before we detail each component in this section, we provide the preliminary background.


We construct a homogeneous graph $\C{G}=(\C{V},\C{E})$ to provide a uniform view of items' multimodality side information, and their relationships. Here, $\C{V}$ denotes the set of nodes (\ie items), and $\C{E}$ denotes the set of edges between them.\footnote{In the context of recommendation, the relationships between items can be defined by their interactions with users, \eg co-purchase or co-click. More details about the construction of item graph are presented in Section~\ref{sss:dataset}.} Each node $h$ has multiple types of side information. We denote the $i$-th modality feature of the node $h$ by $\B{x}_{h}^{i}$, and the number of modality by $m$. 

\begin{algorithm}[t]
  \renewcommand{\algorithmicrequire}{\textbf{Input:}}
  \renewcommand{\algorithmicensure}{\textbf{Output:}}
  \caption{MCNSampling Algorithm}
  \label{alg:mcnsamping}
  \begin{algorithmic}[1]
    \REQUIRE Graph $\C{G}$, batch of nodes $\C{B}$, sampling depth $K$, sampling size $\{n_k\}_{k=1}^{K}$, number of contextual neighbors $S$;
    \ENSURE Sampled contextual neighbors $\C{C}_B$;
    \STATE $\C{C}_B\leftarrow [~]$;
    \FOR{$h \in \C{B}$}
    \STATE $\C{S}_0 \leftarrow [h]$, $\C{S}_1, \C{S}_2, \cdots \C{S}_K \leftarrow [~]$, $s_t \leftarrow 0~\forall t \in \C{V}\setminus h$;
    \FOR{$k=1, 2, \cdots, K$}
        \FOR{$t \in \C{S}_{k-1}$}
            \STATE Sample $n_k$ nodes with replacement from $\C{N}_t$ and append them to $\C{S}_k$;
        \ENDFOR
        \STATE Count the frequency $f_t^k$ of each distinct node $t$ in $\C{S}_k$ and update its score $s_{t} \leftarrow s_{t} + f^k_{t}*(K-k+1)$;
    \ENDFOR
    \STATE Sort the nodes in $\C{V}\setminus h$ according to the importance scores in descending order;
    \STATE Choose $S$ top-ranked nodes as the contextual neighbors $\C{C}_h$ of $h$ and append it to $\C{C}_B$;
    \ENDFOR
    \RETURN{$\C{C}_B$}
  \end{algorithmic}
\end{algorithm}

In $\C{G}$, we denote the one-hop neighbors of $h$ by $\C{N}_h$, and use $\omega_{ht}$ to denote the weight of the edge between two nodes $h$ and $t$, where $\omega_{ht} >0$.
For a node $h$, we use $\C{C}_h$  to denote its contextual neighbors selected by a sampling algorithm, \eg the MCNSampling algorithm.
Given the item graph $\C{G}$ and the contextual neighbors of each node, PMGT aims to obtain the node representations that can capture the multimodality information of nodes and the graph structure. Then, the learned node representations can be applied in downstream tasks directly or with adjustments such as fine-tuning.

\subsection{Contextual Neighbors Sampling}

For each node $h$, there exist some relevant nodes in the graph that may help enrich its representation. These relevant nodes are referred as the \textit{contextual neighbors} of $h$. To efficiently select contextual neighbors for a batch of nodes during the training of PMGT, we develop a sampling algorithm named  MCNSampling.

MCNSampling iteratively samples a list of  nodes for a target node $h$ with a predefined sampling depth $K$.
Let $\C{S}_h^{k-1}$ denote the bag of nodes sampled at the $(k-1)$-th step. For each node $t$ in $\C{S}_h^{k-1}$, we randomly sample $n_k$ nodes with replacement from $t$'s one-hop neighbors $\C{N}_t$ at the $k$-th step. The probability that a node $t'\in \C{N}_t$ being sampled is proportional to the weight $\omega_{tt'}$ of the edge between nodes $t$ and $t'$. Note that a node may appear multiple times in $\C{S}_h^{k-1}$. In the MCNSampling algorithm, we treat all node instances in $\C{S}_h^{k-1}$ as ``\emph{different nodes}'' and perform the sampling procedure.

In our sampling algorithm, we select contextual neighbors by considering 1) the sampled frequency of a node, and 2) the number of sampling steps between the target node $h$ and a sampled node in the sampling process. For every node $t \in \C{V} \setminus h$, we empirically define its importance to the target node $h$ at the $k$-th sampling step ($k\leq K$) as follows,
\begin{equation}
s_{t}^k = f_{t}^k \times (K - k + 1),
\end{equation}
where $f_{t}^k$ denotes the number of times $t$ appearing in $\C{S}_h^k$.
That is, a node $t$ is considered more relevant to the target node $h$, if $t$ is sampled more frequently and it has a smaller sampling steps to $h$. The final importance score of a node $t$ to $h$ is defined as follows,
\begin{equation}
  s_t=\sum_{k=1}^{K}s_t^k.
\end{equation}
Then, we sort all nodes in $\C{V} \setminus h$ according to their importance scores in descending order, and choose the $S$ top-ranked nodes as the sampled contextual neighbors of $h$. The details of the MCNSamping algorithm are summarized in Algorithm~\ref{alg:mcnsamping}.

\subsection{Node Embedding Initialization}

After the neighborhood sampling, we concatenate the target node $h$ and its ordered contextual neighbors $\C{C}_h$, denoted by $\C{I}_h = [h, h_1, h_2, \cdots, h_S]$.  $h_j$ is the $j$-th node in $\C{C}_h$, and $1 \leq j\leq S$. For each node $t \in \C{I}_h$, we apply the attention mechanism to obtain its multimodal representation $\B{M}_t$ as follows,
\begin{gather}
	\B{X}_t^{i} = \B{x}_{t}^{i}\B{W}_{M}^i  + \B{b}_M^i, \nonumber\\
	\B{X}_{t} = \B{X}_t^{1} \oplus \B{X}_t^{2} \oplus \cdots \oplus \B{X}_t^{m},\nonumber\\
	\B{\alpha}_t = \mbox{softmax}\bigg[\mbox{tanh}(\B{X}_{t})\B{W}_s + \B{b}_s\bigg],  \nonumber\\
	\B{M}_{t} = \sum_{i}^{m}\alpha_{t}^i \B{X}_t^{i},
	\label{eq:multimodal-feat}
\end{gather}
where $\B{W}_M^i \in \R{R}^{d_{i}\times d_0}$ and $\B{b}_M^i \in \R{R}^{1 \times d_0}$ denote weight matrix and bias term for the $i$-th modality, $\B{W}_s \in \R{R}^{(md_0) \times m}$ and $\B{b}_s \in \R{R}^{1 \times m}$ denote weight matrix and bias term for attention mechanism. $\oplus$ is the concatenation operation. That is, the multimodality side information of each item is concatenated to contribute to the comprehensive representation learning.   $\alpha_t^i$ denotes the $i$-th element of $\B{\alpha}_t$.

The position of a node in the input list $\C{I}_h$ reflects its importance to the target node $h$.
Thus, we argue that the order of nodes in $\C{I}_h$ is important in learning  node representations. The following position-id embedding is used to identify the node order information of an input list,
\begin{equation}
	\B{P}_t = \mbox{\textbf{P-Embedding}}\bigg[p(t)\bigg],
\end{equation}
where $p(t)$ denotes the position id of  node $t$ in $\C{I}_h$, $\B{P}_t \in \R{R}^{1\times d_0}$ denotes the position-based embedding for $t$.

Our  main objective is to obtain the representation of the target node $h$. Intuitively, the target node and its contextual neighbors should play different roles in the pre-training. To identify the role differences, we add the following role-based embedding to each node $t \in \C{I}_h$,
\begin{equation}
\B{R}_t =  \mbox{\textbf{R-Embedding}}\bigg[r(t)\bigg],
\end{equation}
where $r(t)$ and $\B{R}_t \in \R{R}^{1\times d_0}$ denote the role label and role-based embedding of the node $t$, respectively. In practice,  we set the role label of the target node as ``\texttt{Target}'' and the role labels of the contextual neighbors as ``\texttt{Context}''.


Based on the embeddings stated above, we aggregate them together to define the initial input embedding for a node $t \in\C{I}_h$ as follows,
\begin{equation}
\B{H}_t^0 = \mbox{\textbf{Aggregate}}\bigg(\B{M}_t, \B{P}_t, \B{R}_t\bigg).
\end{equation}
In this work, we simply define the $\mbox{\textbf{Aggregate}}(\cdot)$ function as  vector summation. The initial input embeddings for the nodes in the input list $\C{I}_h$ can be stacked into a matrix $\B{H}^0=[\B{H}_0^0; \B{H}_1^0; \cdots; \B{H}_S^0] \in \R{R}^{(S+1) \times d_0}$, where $\B{H}_0^0$ corresponds to the target node $h$.

\subsection{Transformer-based Graph Encoder}

We use the Transformer framework~\cite{vaswani2017attention}  to model the mutual influences between a node and its contextual neighbors.  
Given the node representations $\B{H}^{\ell-1}$ at the $(\ell-1)$-th layer, the output at the $\ell$-th layer of original Transformer model is defined as follows,
\begin{gather}
	\B{H}^\ell = \mbox{FFN}\bigg[\mbox{softmax}(\frac{\B{Q}\B{K}^\top}{\sqrt{d_h}})\B{V}\bigg], \nonumber \\
	\B{Q} = \B{H}^{\ell-1}\B{W}_Q^{\ell}, \B{K} = \B{H}^{\ell-1}\B{W}_K^{\ell}, \B{V} = \B{H}^{\ell-1}\B{W}_V^{\ell},
\end{gather}
where $\B{W}_Q^{\ell},\B{W}_K^{\ell},\B{W}_V^{\ell} \in \R{R}^{d_0 \times d_0}$ denote the weight matrices, $\mbox{FFN}(\cdot)$ is the feed forward network. Here, we omit the residual network in the formula for convenience.

\begin{algorithm}[t]
  \renewcommand{\algorithmicrequire}{\textbf{Input:}}
  \renewcommand{\algorithmicensure}{\textbf{Output:}}
  \caption{PMGT Optimization Algorithm}
  \label{alg:PMGT}
  \begin{algorithmic}[1]
    \REQUIRE Item Graph $\C{G}$, multimodal features $\{\B{x}^i\}_{i=1}^m$ of items
    \ENSURE Parameters $\B{\Theta}$ of PMGT and pre-trained item representations
    \STATE Randomly initialize all parameters;
    \FOR{$iter=1, 2, \cdots, max\_iter$}
    \STATE Sample a batch of nodes $\C{B}$ from $\C{G}$;
    \STATE Sample the contextual neighbors $\C{C}_B$ by MCNSampling Algorithm;
    \STATE Initilize the node embeddings of $\C{C}_B$ with multimodal features;
    \STATE Compute gradients of objective function with respect to $\B{\Theta}$ by back-propagation, based on nodes in $\C{B}$;
    \STATE Update $\B{\Theta}$ by gradient descent algorithm(\ie Adam) with learning rate $\eta$;
    \ENDFOR
    \STATE Obtain the item representations pre-trained by PMGT;
    \RETURN{$\B{\Theta}$} and the pre-trained item representations;
  \end{algorithmic}
\end{algorithm}

For the target node $h$, there may exist some sampled nodes in $\C{C}_h$, whose representations are similar to the representation of $h$. 
Assume that all the sampled contextual neighbors are relevant to the target node. We hope the proposed model can capture the diversity of the sampled contextual neighbors, by concentrating on the nodes that are relevant but not very similar to the target node. To achieve this objective, we design a diversity-promoting attention mechanism and include it into the attention network of Transformer,
\begin{gather}
\B{S} = \B{H}^{\ell-1}\B{W}_S^\ell, \nonumber\\
\B{U}_1 = \mbox{softmax}\bigg(\B{E} - \frac{\B{S}\B{S}^\top}{||\B{S}||_2 ||\B{S}||_2^\top} + \B{I}\bigg), \nonumber\\
\B{U}_2 = \mbox{softmax}(\frac{\B{Q}\B{K}^\top}{\sqrt{d_h}}), \nonumber \\
\B{H}^\ell = \mbox{FFN}\bigg[(\beta\B{U}_1 + (1 - \beta)\B{U}_2)\B{V}\bigg],
\label{eq:divattention}
\end{gather}
where $\B{W}_S^l \in \R{R}^{d_0 \times d_0}$ is the weight matrix, $\B{E} \in \R{R}^{(S+1) \times (S+1)} $ is a matrix where all its elements are 1, $||\B{S}||_2 \in \R{R}^{(S+1) \times 1}$ denotes the $\ell_2$ row norm of $\B{S}$, and $\B{I} \in \R{R}^{(S+1) \times (S+1)}$ denotes the identity matrix. Note that the larger the similarity between two different nodes, the smaller the attention weight between them in $\B{U}_1$. The objective of adding $\B{I}$ in the definition of $\B{U}_1$ is to include the node's self information. $\beta$ is a constant ($0\leq \beta \leq 1$) balancing the contributions of the two attention weights.

After obtaining the output $\B{H}^L$ at the last layer of the encoder, we obtain $\B{H}^L_0$ as the representation of target node $h$, denoted as $\B{h}$ for simplicity. Then, $\B{H}^L$ will be used in the following pre-training tasks.

\subsection{Model Optimization}

The proposed PMGT model is pre-trained with the following two objectives: 1) graph structure reconstruction, and 2) masked node feature reconstruction. To ensure the learned node representations can capture the graph structure, 
we define the following loss function~\cite{hamilton2017inductive},
\begin{align}
\C{L}_{edge} = & \frac{1}{|\C{V}|}\sum_{h \in \C{V}}\frac{1}{|\C{N}_h|}\sum_{t \in \C{N}_h}\bigg[-\log\big(\sigma(\frac{\B{h}^T\B{t}}{||\B{h}||_2||\B{t}||_2})\big) \notag \\
& - Q*E_{t_n \sim P_n(t)}\log\big(\sigma(-\frac{\B{h}^T\B{t}_n}{||\B{h}||_2||\B{t_n}||_2})\big)\bigg],
\label{eq:edgeloss}
\end{align}
where $\sigma (\cdot)$ is the sigmoid function, $P_n$ and $Q$ denote the negative sampling distribution and the number of negative samples, respectively.

The node feature reconstruction task focuses on capturing the multimodal features in the learned node representations. Previous methods, \eg GRAPH-BERT~\cite{zhang2020graph}, design an attribute reconstruction task without masking operations. Thus, the models' abilities in aggregating the features of different nodes may be limited. In this work, we design a masked node feature reconstruction task, which aims to reconstruct the features of masked nodes by other non-masked nodes in $\C{I}_h$. As the representation of the target node $h$ is needed to reconstruct the graph structure in Eq.~\eqref{eq:edgeloss}, we do not apply the masking operation to the target node $h$. Following~\cite{devlin2018bert}, we randomly choose 20\% of nodes in the list $\C{I}_h \backslash h$ for masking. If the node $t$ is chosen, we replace $t$ with: 1) the [\texttt{Mask}] node 80\% of the time, (2) a random node 10\% of the time, and (3) the unchanged node $t$ 10\% of the time. Then, the masked item list will be input to the model, and the output $\B{H}^L$ will be used to reconstruct the multimodal features of the masked nodes. We set the input features of the [\texttt{Mask}] node to $\B{0}$, and define the feature reconstruction loss as follows,
\begin{equation}
\C{L}_{feature} = \frac{1}{|\C{V}|}\sum_{h\in \C{V}}\frac{1}{|\C{M}_h|}\sum_{t\in \C{M}_h}\sum_{i}^{m}\bigg\|\B{H}^L_t\B{W}_r^i -\B{x}_t^{i}\bigg\|_2^2,
\end{equation}
where $\C{M}_h$ denotes the set of masked nodes in $\C{I}_h$, $\B{H}_t^L$ denotes the representation of $t$ in $\B{H}^L$, and $\B{W}_r^i$ is the weight matrix for the $i$-th modality information reconstruction.

The model parameters of PMGT can be learned by minimizing the combined  objective function,
\begin{equation}\label{eq:objectives}
  \C{L}_{edge} + \C{L}_{feature}
\end{equation}
The entire framework can be effectively trained by the end-to-end backpropagation algorithm. To make the training of the model more stable, a mini-batch of nodes are randomly sampled to update the model. The details of the training procedure for the proposed PMGT model are summarized in Algorithm~\ref{alg:PMGT}. When applying the pre-trained PMGT model in downstream tasks, the learned node representations can be either fed into the new tasks directly or with necessary adjustment (\eg fine-tuning).



\section{Experiments on Recommendation Tasks}
\label{sec:recexperiments}

We conduct experiments on two real datasets: the Amazon review dataset~\cite{ni2019justifying} and Movielens-20M dataset\footnote{https://grouplens.org/datasets/movielens/20m/}. We firstly use two downstream tasks, \ie item recommendation and CTR prediction to evaluate the effectiveness of the proposed PMGT model.
\subsection{Experimental Settings}
\subsubsection{Experimental Datasets}
\label{sss:dataset}
\begin{table}
	\centering
	\caption{Statistics of the experimental datasets for item recommendation and CTR prediction tasks.}
	\label{tab:tab1}
	\begin{tabular}{l|rrr|rr}
\hline
\multirow{2}{*}{Datasets} & \multicolumn{3}{c|}{Data for Downstream Tasks} & \multicolumn{2}{c}{Item Graph} \\ \cline{2-6}
                         & \# Users  & \# Items & \# Interact. & \# Nodes       & \# Edges       \\ \hline
VG & 4,525 & 3,921 & 27,780 & 5,032 &83,981 \\
		TG & 31,109  & 13,870  & 182,744& 17,388 &232,720   \\
		THI & 20,082 & 11,170& 109,717& 15,619 &178,834 \\
		ML & 27,715 & 4,253 & 2,179,386& 4,271 &249,498 \\ \hline
\end{tabular}
\end{table}

\begin{table*}
	\caption{Performances of item recommendation (REC) and CTR prediction by using different pre-training methods. Best results are in boldface and second best underlined.  
	}
	\centering
	\small
	\label{tab:tab2}
	\begin{tabular}{l|l|cccccccc}
		\hline
		Datasets & Metrics & Random & DeepWalk & LINE & TransAE & GraphSAGE & GRAPH-BERT & GPT-GNN & PMGT\\\hline
		\multirow{5}{*}{VG}
		& REC-R@10 & 0.1994 & 0.2236 & 0.2234 & 0.1989 & 0.2052 & \underline{0.2380} & 0.2251 & \textbf{0.2480} \\
		& REC-N@10 & 0.1278	& 0.1441 & 0.1420 & 0.1217 & 0.1301 & \underline{0.1491} & 0.1343 & \textbf{0.1625} \\
		& REC-R@20 & 0.2742 & 0.3179 & 0.3169 & 0.2903 & 0.2821 & \underline{0.3330} & 0.3269 & \textbf{0.3405}\\
		& REC-N@20 & 0.1494 & 0.1711 & 0.1690 & 0.1480 & 0.1522 & \underline{0.1767} & 0.1636 & \textbf{0.1890}\\
		& CTR-AUC & 0.7311 & 0.768 & 0.7762 & 0.7675 & 0.7674 & 0.7746 & \underline{0.7839} & \textbf{0.7990}\\
		\hline \hline
		\multirow{5}{*}{TG}
		& REC-R@10 & 0.2147 & 0.2787 & 0.2805 & 0.2137 & 0.2391 & \underline{0.2858} & 0.2598 & \textbf{0.3032}\\
		& REC-N@10 & 0.138 & 0.1847 & 0.1857 & 0.1358 & 0.1514 & \underline{0.1942} & 0.1671 & \textbf{0.2056} \\
		& REC-R@20 & 0.3068 & 0.3807 & \underline{0.3873} & 0.3051 & 0.3352 & 0.3808 & 0.3608 & \textbf{0.4030}\\
		& REC-N@20 & 0.1644 & 0.2141 & 0.2162 & 0.1620 & 0.1790 & \underline{0.2215} & 0.1962 & \textbf{0.2342}\\
		& CTR-AUC & 0.8047 & 0.8289 & 0.8326 & 0.8214 & 0.8266 & 0.8322 & \underline{0.8328} & \textbf{0.8370}\\
		\hline\hline
		\multirow{5}{*}{THI}
		& REC-R@10 & 0.1957 & \underline{0.2043} & 0.1626 & 0.1425 & 0.1921 & 0.2023 & 0.1680 & \textbf{0.2358}\\
		& REC-N@10 & 0.136 & 0.1399 & 0.0996 & 0.0861 & 0.1320 & \underline{0.1462} & 0.1047 & \textbf{0.1707} \\
		& REC-R@20 & 0.2555 & \underline{0.2756} & 0.2403 & 0.2191 & 0.2577 & 0.2627 & 0.2451 & \textbf{0.3025}\\
		& REC-N@20 & 0.1529 & 0.1598 & 0.1214 & 0.1077 & 0.1505 & \underline{0.1632} & 0.1264 & \textbf{0.1895}\\
		& CTR-AUC & 0.7652 & 0.7850 & \underline{0.7896} & 0.7815 & 0.7643 & 0.7878 & 0.7817 & \textbf{0.7933}  \\
		\hline\hline
		\multirow{5}{*}{ML}
		& REC-R@10 & 0.3104	& 0.3144 & 0.3112 & 0.3096 & 0.3139 & 0.3136 & \textbf{0.3162} & \underline{0.3161}\\
        & REC-N@10 & 0.4714 & 0.4793 & 0.4763 & 0.4703 & 0.4769 & 0.4781 & \textbf{0.4823} & \underline{0.4814}\\
		& REC-R@20 & 0.4556 & 0.4618 & 0.4589 & 0.4549 & 0.4606 & 0.4581 & \underline{0.4628} & \textbf{0.4640}\\
		& REC-N@20 & 0.4829 & \underline{0.4911} & 0.4881 & 0.4822 & 0.4888 & 0.4883 & \textbf{0.4928} & \textbf{0.4928}\\
		& CTR-AUC & 0.9109 & \textbf{0.9218} & 0.9200 & 0.9194 & 0.9145 & 0.9184 & 0.9191 & \underline{0.9205}  \\
		\hline
	\end{tabular}
\end{table*}
\textbf{Amazon Datasets.} We choose the following 5-score review subsets of the Amazon review dataset for studying the item recommendation and CTR prediction tasks, \ie ``Video Games'', ``Toys and Games'', and ``Tools and Home Improvement'' (respectively denoted by VG, TG, and THI). The metadata of a product includes its text description and the URL of its image\footnote{https://nijianmo.github.io/amazon/index.html}, which are used to extract the textual and visual features of the product respectively. In the experiments, we use the rating data generated before 2015-01-01 for building the product graph, and the rating data generated since 2015-01-01 for studying the performances of the item recommendation and CTR prediction tasks. In these two downstream tasks, we convert all the observed review ratings to be positive interactions and filter out the products that are not included in the product graph.

Moreover, we build the product graph based on users' review behaviors. Let $r_{ht}$ denotes the number of users who have commonly reviewed the two products $h$ and $t$. If $r_{ht} \geq 3$, we connect the two products $h$ and $t$ in the graph. In the graph, there inevitably exist some popular nodes with large degrees. This usually causes that the popular nodes with large degrees are more likely to be sampled in the contextual neighbors sampling procedure. To alleviate this problem, we define the edge weight based on the vertex degrees of an edge. Empirically, we define the weight $\omega_{ht}$ of the edge $e_{ht}$ between the nodes $h$ and $t$ as follows,
\begin{equation}
	\omega_{ht} = \frac{\log(r_{ht}) + 1}{\log(\sqrt{d_h\ast d_t}) + 1},
\label{eq:edgeweight}
\end{equation}
where $d_h$ and $d_t$ denote the degrees of the nodes (\ie products) $h$ and $t$ in the graph, respectively. The operation of $\log(\cdot)$  alleviates the problem of large variance of edge weights in the product graph. By inversely proportional to the degrees of nodes $h$ ad $t$, we  reduce the weights of popular nodes.


\noindent \textbf{Movielens-20M Dataset.}
For the Movielens-20M dataset (denoted by ML), we construct the movie graph based on the tags of movies. The tags of a movie are obtained from the MovieLens Tag Genome Dataset\footnote{https://grouplens.org/datasets/movielens/tag-genome/}, which includes 11 million computed tag-movie relevance scores from a pool of 1,100 tags applied to 10,000 movies. For each movie, we only keep the tags with relevance scores larger than 0.9. Then, we use $r_{ht}$ to denote the number of common tags  two movies $h$ and $t$ have. If $r_{ht} \geq 3$, we construct an edge between the two movies $h$ and $t$, and the weight $\omega_{ht}$ of the edge $e_{ht}$ between two nodes $h$ and $t$ is defined following Eq.~\eqref{eq:edgeweight}. We collect the movie trailers from Youtube\footnote{https://www.youtube.com/} and extract keyframes for each movie trailer. These keyframes are used as the movie images for extracting the visual modality of a movie. The movie descriptions are collected from TMDB\footnote{https://www.themoviedb.org/}. In addition, the rating data generated since 2008-01-01 are used to evaluate the performances of item recommendation and CTR prediction tasks, where we keep ratings larger than 3 as positive interactions.

\begin{figure*}[ht]
	\centering
	\includegraphics[width=\textwidth]{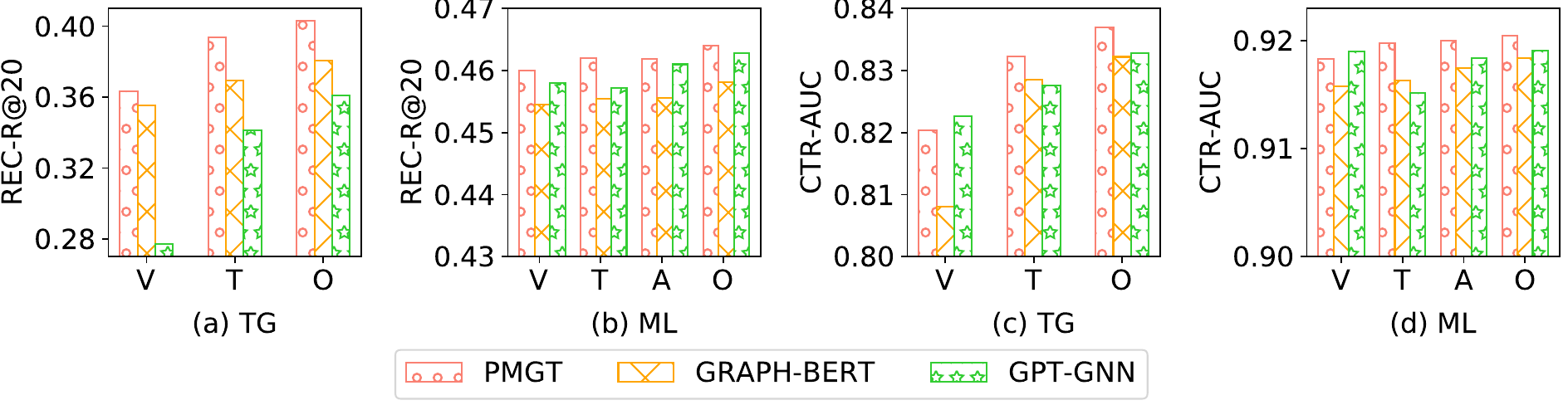}
	\caption{Performances of PMGT, GRAPH-BERT and GPT-GNN considering different modality information on TG and ML datasets. V, T, A denotes the visual, textual, and acoustic modality information respectively. O denotes  original models considering all the modality information. }
	\label{fig:ablationmodality}
\end{figure*}
\begin{figure*}[ht]
	\centering
	\includegraphics[width=\textwidth]{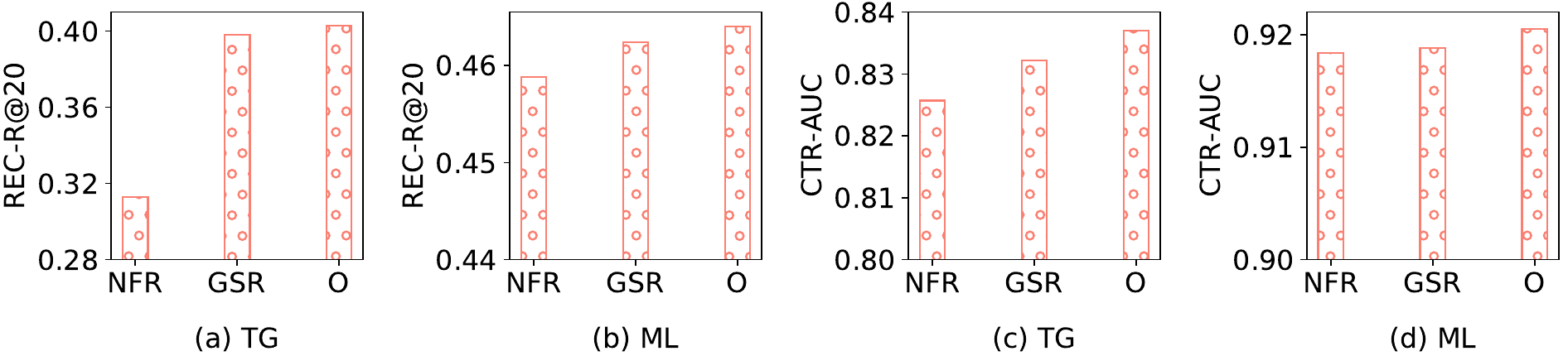}
	\caption{Performances of PMGT considering different graph reconstruction tasks on TG and ML datasets. NFR and GSR denote PMGT only considering the node feature reconstruction task and graph structure reconstruction task, respectively. O denotes the original PMGT considering both tasks.}
	\label{fig:ablationtask}
\end{figure*}

\noindent\textbf{Multimodal Feature Extraction.} In the experiments, we use the pre-trained Inception-v4 network to extract the visual features of each image. Then, we average the visual features of all the images of an item (\ie product or movie) to obtain its visual modality. For the text description of an item, we utilize the pre-trained BERT model to extract the features of each sentence. Then, we average the features of all the sentences in the description of an item to obtain its textual modality. Empirically, we set the length of the sentence to 128. For the ML dataset, we separate the audio track from the movie trailer with FFmpeg\footnote{http://ffmpeg.org/} and adopt VGGish~\cite{hershey2017cnn} to obtain the acoustic modality of the movie. The dimensionality of the visual, textual, and acoustic modalities are 1,536, 768, and 128, respectively. Table~\ref{tab:tab1} summarizes the statistics of these experimental datasets used for item recommendation and CTR prediction tasks.

\subsubsection{Setup and Metrics}
\label{sss:setup}

After the data pre-processing in Section~\ref{sss:dataset}, an item multimodal graph $\C{G}$ is built, and a set of user-item interactions $\C{D}_{down}$ is prepared for studying downstream tasks.

In the pre-training task, we randomly keep 90\% of nodes and their relationships in $\C{G}$ to pre-train the item representations. The remaining 10\% of nodes (denoted by $\C{V}_{val}$) are used to construct the validation data for choosing the hyper-parameters of different pre-training models. For each node $t$ in $\C{V}_{val}$, we randomly sample a node $t_{+}$ from its one-hop neighbors in $\C{G}$, and use the pair $(t, t_{+})$ as a positive example in validation data. And we also randomly sample a node $t_{-}$ that is not connected with $t$ in $\C{G}$, and use the pair $(t, t_{-})$ as a negative example in validation data. The model parameters are chosen based on the AUC (\ie Area under the ROC Curve) computed on all the validation data.

For downstream tasks, we choose Neural Collaborative Filtering (NCF)~\cite{he2017neural} as the base model for item recommendation task, and Deep \& Cross Network (DCN)~\cite{wang2017deep} as the base model for CTR prediction task.
\begin{itemize}
    \item For item recommendation task, we randomly select 80\% of the interactions in $\C{D}_{down}$ as training data to update the NCF model, and the remaining 20\% of interactions in $\C{D}_{down}$ are used for testing. Moreover, we also randomly hold out 10\% of the training data for tuning the hyper-parameters of NCF. In the training of NCF, for each positive interaction pair, we randomly sample one item that has no interactions with the user as negative feedback to update the model. The item recommendation performances achieved by NCF are evaluated by Recall@10, Recall@20, NDCG@10, and NDCG@20 (respectively denote by REC-R@10, REC-R@20, REC-N@10, and REC-N@20). To improve the evaluation efficiency, we randomly sample 1000 items that the testing user has not interacted with to compute the recall and NDCG.

    \item For CTR prediction task, we use $\C{D}_{down}$ to simulate the CTR data. For each possible user-item  pair in $\C{D}_{down}$, we randomly sample 5 items that have no interactions with the user to construct the negative samples of the CTR data. All the interaction pairs in $\C{D}_{down}$ are used as the positive examples of the CTR data. Then, we randomly sample 80\% of the CTR data to train the DCN model, and use the remaining 20\% of CTR data to testing the CTR prediction performances. Moreover, 10\% of the training data are also held out for tuning the hyper-parameters of DCN. The CTR prediction performances are evaluated by AUC (denoted by CTR-AUC).

\end{itemize}

\begin{figure*}
     \centering
     \begin{subfigure}[b]{0.32\textwidth}
         \centering
         \includegraphics[width=\textwidth]{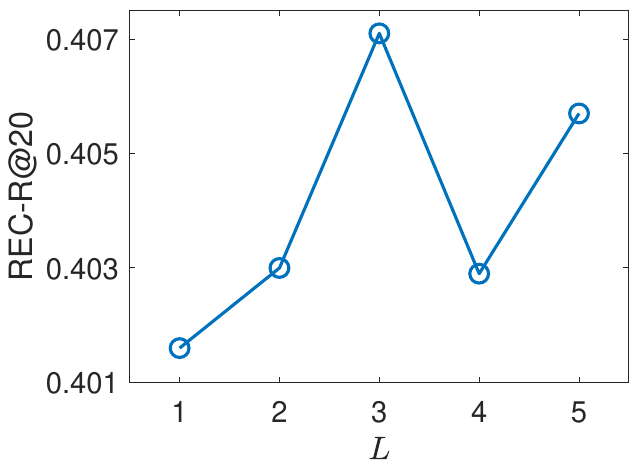}
         \caption{}
         \label{fig:layers1}
     \end{subfigure}
     \hfill
     \begin{subfigure}[b]{0.32\textwidth}
         \centering
         \includegraphics[width=\textwidth]{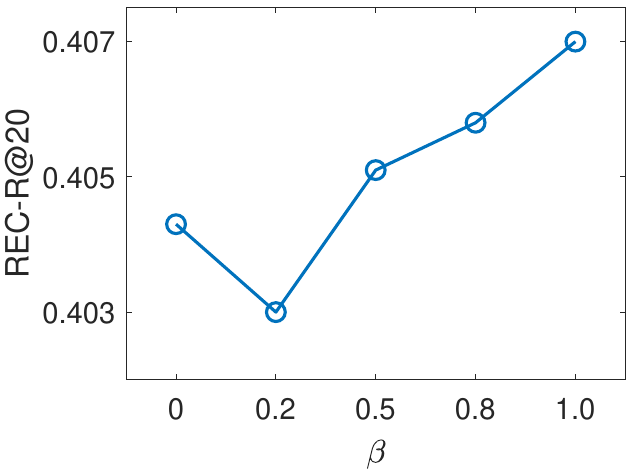}
         \caption{}
         \label{fig:beta1}
     \end{subfigure}
     \hfill
     \begin{subfigure}[b]{0.32\textwidth}
         \centering
         \includegraphics[width=\textwidth]{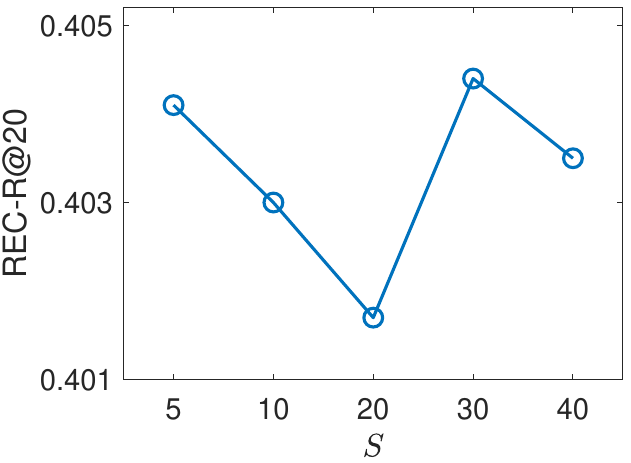}
         \caption{}
         \label{fig:size1}
     \end{subfigure}
    \caption{The recommendation performance trend of NCF based on item representations pre-trained by PMGT with respect to different settings of $L$, $\beta$, and $S$ on TG dataset.}
    \label{fig:HyperPara1}
\end{figure*}
\begin{figure*}
     \centering
     \begin{subfigure}[b]{0.32\textwidth}
         \centering
         \includegraphics[width=\textwidth]{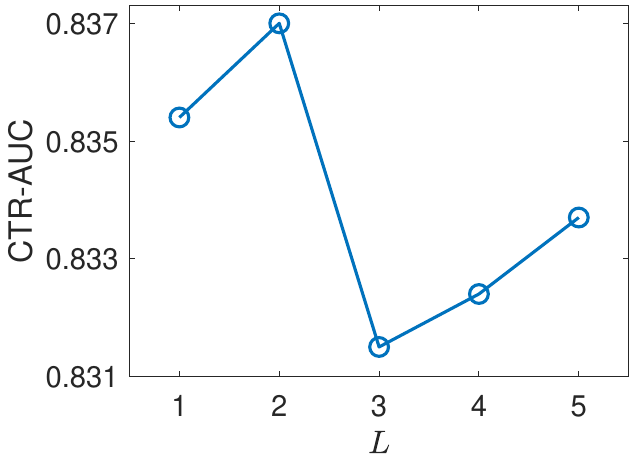}
         \caption{}
         \label{fig:layers2}
     \end{subfigure}
     \hfill
     \begin{subfigure}[b]{0.32\textwidth}
         \centering
         \includegraphics[width=\textwidth]{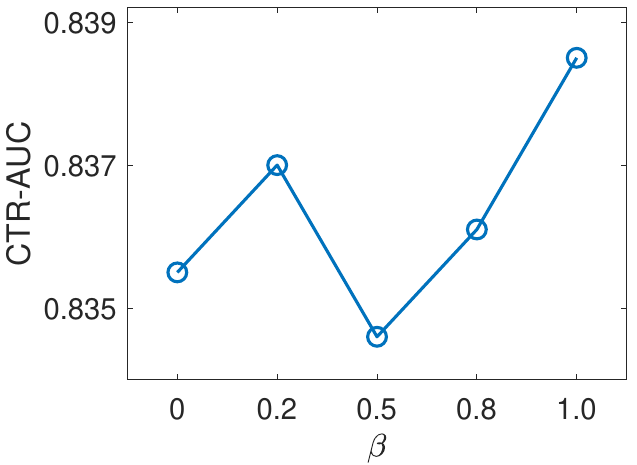}
         \caption{}
         \label{fig:beta2}
     \end{subfigure}
     \hfill
     \begin{subfigure}[b]{0.32\textwidth}
         \centering
         \includegraphics[width=\textwidth]{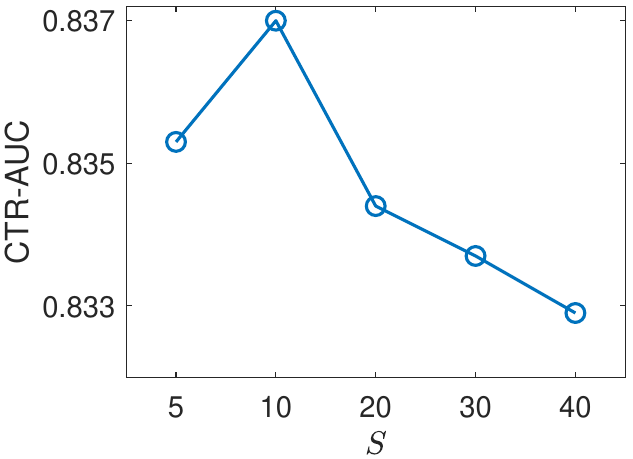}
         \caption{}
         \label{fig:size2}
     \end{subfigure}
    \caption{The CTR prediction performance trend of DCN based on item representations pre-trained by PMGT with respect to different settings of $L$, $\beta$, and $S$ on TG dataset.}
    \label{fig:HyperPara2}
\end{figure*}

\subsubsection{Baseline Methods}

We compare the proposed PMGT model with the following pre-training methods:

\begin{itemize}
  \item \textbf{Random}: The item embeddings in the downstream tasks are randomly initialized.

  \item \textbf{DeepWalk}~\cite{perozzi2014deepwalk}: This method learns node representations by sampling a large number of paths in the graph and maximizing the average logarithmic probability of all vertex context pairs in sampled paths.
  \item \textbf{LINE}~\cite{tang2015line}: This graph embedding method is trained to preserve the first- and second-order proximities of nodes in the graph.
  \item \textbf{GraphSAGE}~\cite{hamilton2017inductive}: This GNN model forces connected nodes to have similar embeddings by aggregating the information of neighboring nodes.
  \item \textbf{TransAE}~\cite{wang2019transae}: This method combines a multimodal encoder and the TransE~\cite{antoine2013trane} model to learn the node representations.
  \item \textbf{GRAPH-BERT}~\cite{zhang2020graph}: This method applies Transformer to aggregate neighbors' information without masking operations on the nodes.
  \item \textbf{GPT-GNN}~\cite{hu2020gpt}: This method 
employs the attribute generation and edge generation tasks to pre-train the GNN model.
\end{itemize}

For a fair comparison, we use the same multimodal representation in Eq.~\eqref{eq:multimodal-feat} as the inputs for all pre-training methods.

\subsubsection{Implementation Details}
For the pre-training and downstream tasks, we set the dimensionality of latent space $d_0$ to 128. In the experiments, we empirically set the sampling depth $K$ to 3, and the sampling sizes $n_1, n_2, n_3$ to 16, 8, 4 respectively. The number of contextual neighbors $S$ is selected from \{5, 10, 20, 30, 40\}. The number of transformer layers $L$ is chosen from \{1, 2, 3, 4, 5\}. The weight of diversity-promoting attention $\beta$ is selected from \{0, 0.2, 0.5, 0.8, 1.0\}. We implement PMGT based on TensorFlow~\cite{abadi2016tensorflow} and AliGraph~\cite{zhu2019aligraph} frameworks. Adam~\cite{kingma2014adam} is used as the optimizer for learning model parameters, and the learning rate is chosen from \{$10^{-4}, 10^{-3}, 10^{-2}$\}.

\subsection{Performance Comparison}

After pre-training on the item graph, we use the pre-trained item representations to initialize the item embeddings in the downstream tasks. Then, we train the NCF and DCN models and fine-tune the item embeddings based on the user-item interaction data. Table~\ref{tab:tab2} summarizes the performances of NCF and DCN initialized with item representations pre-trained by different methods. We make the following observations. Compared with the random initialization, initializing the base models with pre-trained item representations usually achieves better item recommendation and CTR prediction performances. This demonstrates the pre-training strategies can benefit downstream tasks in recommendation scenarios. The deep pre-training methods GRAPH-BERT, GPT-GNN, and PMGT usually outperform other shallow pre-training methods, by employing GNN to aggregate the neighbor information,  and using node feature reconstruction and graph structure reconstruction tasks to pre-train the model. Moreover, PMGT usually achieves the best item recommendation and CTR prediction performances on all datasets. This demonstrates the effectiveness of PMGT in exploiting the item graph structure and item features.
In addition, we can also note that PMGT achieves smaller improvements on the ML dataset. One potential reason is that the interaction data of the ML dataset are denser. Thus, with the random initialization, the base models can learn sufficiently good item representations for downstream tasks.

%
%


\begin{figure*}
     \centering
     \begin{subfigure}[b]{0.4\textwidth}
         \centering
         \includegraphics[width=\textwidth]{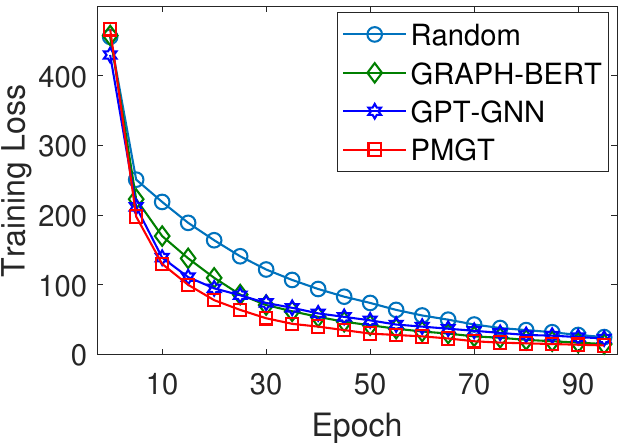}
         \caption{NCF}
         \label{fig:ncfloss}
     \end{subfigure}
     \begin{subfigure}[b]{0.4\textwidth}
         \centering
         \includegraphics[width=\textwidth]{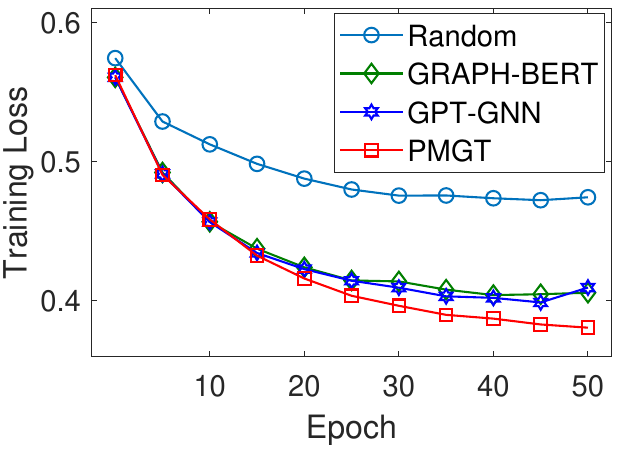}
         \caption{DCN}
         \label{fig:dcnloss}
     \end{subfigure}
    	\caption{The convergence speed of the training loss of (a) NCF model and (b) DCN model on the TG dataset.}
    \label{fig:convergence}
\end{figure*}

\subsection{Ablation Study}

We also study the effectiveness of PMGT in exploiting different modality information.
As shown in Figure~\ref{fig:ablationmodality}, we can note that the original methods considering multimodality information usually outperform the variants that only consider single modality information. This observation is as expected. It indicates that representing items with multimodality information can achieve better performance. PMGT is superior to GRAPH-BERT and GPT-GNN with considering single modality information in most scenarios. This observation again demonstrates that PMGT is more effective in capturing different types of modality information than baseline methods.

Moreover, we also study the effectiveness of the two graph reconstruction tasks in learning node representations.
Figure~\ref{fig:ablationtask} summarizes the performances of PMGT variants on the TG and ML datasets. We can note that the original PMGT model with two tasks consistently outperforms the variants using a single task as the pre-training objective. This indicates that both the graph structure reconstruction task and the masked node feature reconstruction task are essential for learning useful node representations to benefit downstream tasks.



\subsection{Parameter Sensitivity Study}

In this section, we study the performances of PMGT with respect to (w.r.t.) different settings of three important hyper-parameters. Firstly, we vary the number of Transformer layers $L$ from 1 to 5. As shown in Figure~\ref{fig:layers1} and Figure~\ref{fig:layers2} , the best item recommendation and CTR prediction performances are achieved by setting $L$ to 3 and 2, respectively. Further stacking more layers does not help improve the performances of downstream tasks. 
Moreover, we vary the weight of the diversity-promoting attention score $\beta$ in \{0, 0.2, 0.5, 0.8, 1.0\}. As shown in Figure~\ref{fig:beta1}, the recommendation accuracy can be improved by considering diversity-promoting attention in the Transformer-based encoder, when $\beta$ is set to 0.5, 0.8, and 1.0. 
This indicates it is important to consider the diversity of contextual neighbors when learning the node representations. Figure~\ref{fig:size1} ad Figure~\ref{fig:size2} summarize the performances of PMGT w.r.t. different settings of the number of contextual neighbors $S$. We observe that PMGT usually achieves good performances by setting $S$ to a small value (\eg 5 and 10). This indicates that a small number of contextual neighbors can capture the important neighborhood information of a node. Further increase of $S$ tends to include noise information, thus may not help improve the model performances.

\begin{table}
	\centering
	\caption{Statistics of the datasets for item classification.}
	\label{tab:itemclassificationdata}
	\begin{tabular}{l|c|c|c} \hline
		& \# Nodes &\# Edges &\# Labels  \\ \hline
		MI & 13,508  &50,067 &19\\ \hline
		IS & 19,106 &83,661 &23 \\\hline
		ML & 4,271 &24,9498 &19 \\ \hline
	\end{tabular}
\end{table}

\begin{table*}
	\caption{Item classification performances achieved by  different pre-training methods. PMG-NP denotes the PMGT without pre-training. Best results are in boldface and second best underlined.
	}
	\centering
	\small
	\label{tab:itemclassification}
	\begin{tabular}{l|l|cccccccc}
		\hline
		Datasets & Metrics & DeepWalk & LINE & TransAE & GraphSAGE & GRAPHBERT& GPT-GNN & PMGT-NP & PMGT \\\hline
		\multirow{3}{*}{MI}
		& R@5 & 0.7910 & 0.9093 & 0.7214 & 0.845 & 0.8500 & \underline{0.9127} & 0.8108 & \textbf{0.9193}\\
		& N@5 & 0.6866 & \underline{0.8037} & 0.6484 & 0.7358 & 0.7363 & 0.7906 & 0.7143 & \textbf{0.8217}\\
		& AUC & 0.8869 & \underline{0.9506} & 0.8684 & 0.9141 & 0.9147 & 0.9464 & 0.8947 & \textbf{0.9552}\\
		\hline \hline
		\multirow{3}{*}{IS}
		& R@5 & 0.8101 & \underline{0.8877} & 0.6026 & 0.8310 & 0.8262 & 0.8652 & 0.8370 & \textbf{0.8936}\\
		& N@5 & 0.6782 & \underline{0.7664} & 0.4476 & 0.7040 & 0.7005 & 0.7442 & 0.7158 & \textbf{0.7790}\\
		& AUC & 0.9167 & \underline{0.952} & 0.8173 & 0.9270 & 0.9252 & 0.9416 & 0.9280 & \textbf{0.9549}\\
		\hline\hline
		\multirow{3}{*}{ML}
		& R@5 & 0.8144 & \underline{0.8251}	& 0.7620 & 0.6495 & 0.7431 & 0.7995 & 0.6313 & \textbf{0.8436}\\
		& N@5 & 0.7324 & \underline{0.7612} & 0.7009 & 0.5833 & 0.6651 & 0.7469 & 0.5542 & \textbf{0.7727}\\
		& AUC & 0.8839 & \underline{0.8846} & 0.8676 & 0.7897 & 0.8469 & 0.8764 & 0.7738 & \textbf{0.8987}  \\
		\hline
	\end{tabular}
\end{table*}
\begin{figure*}
     \centering
     \begin{subfigure}[b]{0.45\textwidth}
         \centering
         \includegraphics[width=\textwidth]{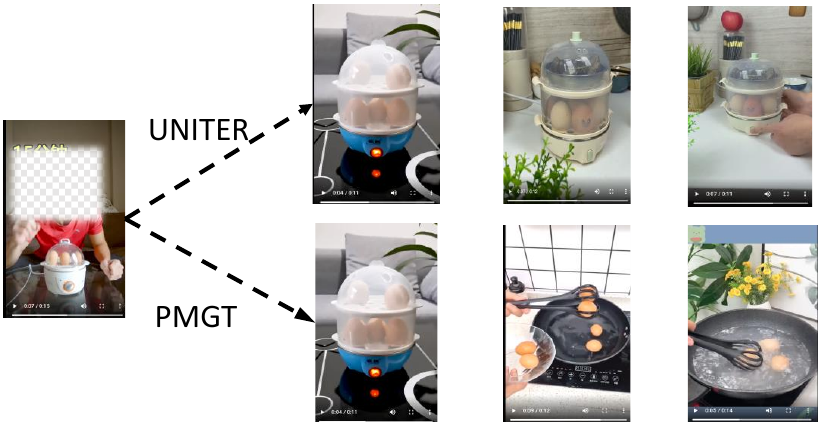}
         \caption{Egg Steamer}
         \label{fig:case1}
     \end{subfigure}
     \hfill
     \begin{subfigure}[b]{0.45\textwidth}
         \centering
         \includegraphics[width=\textwidth]{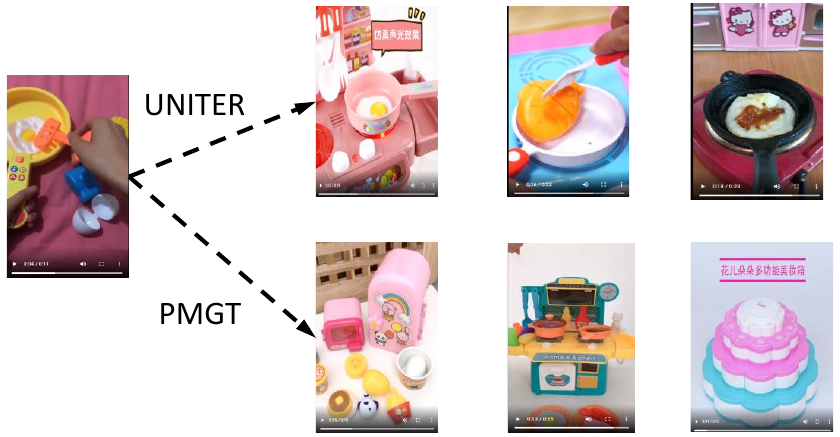}
         \caption{Toy of Pot}
         \label{fig:case2}
     \end{subfigure}
    \caption{Examples of video retrieval based on the video representations pre-trained by PMGT and UNITER.}
    \label{fig:case_study}
\end{figure*}
\subsection{Convergence Speed Study}

Figure~\ref{fig:convergence} shows the convergence speed of the training losses of both NCF and DCN models on the TG dataset. We initialize the item embeddings in the NCF and DCN models by the following strategies: 1) random initialization, 2) initializing using the representations pre-trained by GRAPH-BERT, 3) initializing using the representations pre-trained by GPT-GNN, and 4) initializing using the representations pre-trained by PMGT. We can make the following observations. Compared with the random initialization, initializing item embeddings with the pre-trained representations achieves faster convergence speed. This once again demonstrates the effectiveness of the pre-training strategies. Moreover, both NCF and DCN models achieve the fastest convergence speed by using pre-trained representations by PMGT. The pre-trained PMGT model gives the downstream model a good initial state, thus can help the downstream model achieves faster and earlier convergence.

\section{Experiments on Item Classification}
\label{sec:expitem}

Moreover, we also choose item classification as another downstream task to demonstrate the effectiveness of the proposed pre-training model.

\subsection{Experimental Datasets}
Two 5-core Amazon review subsets, \ie ``Musical Instruments'' and ``Industrial and Scientific'' (respectively denoted by MI and IS), and the ML dataset are used to study the performances of the item classification task. For MI and IS datasets, we use the categories of items as their labels and perform the multi-label classification task. In the experiments, we remove the primary category that almost every product belongs to, for keeping the identification of categories. On the ML dataset, we use the genres of movies as their labels. Finally, there are 19, 23, and 19 labels on the MI, IS, and ML datasets, respectively. The item multimodal graphs of these datasets are built following similar strategies as stated in Section~\ref{sss:dataset}. As the user-item interactions are not required in the item classification task, we use all the review data of the Amazon datasets (\ie MI and IS) to build the item graphs. Table~\ref{tab:itemclassificationdata} summarizes the statistics of these experimental datasets.

\subsection{Setup and Metrics}
In the pre-training stage, we use the same settings as introduced in Section~\ref{sss:setup} to learn parameters of the pre-training models. For item classification, we freeze the pre-trained model and only fine-tune the parameters of the classification model. Following~\cite{hu2020gpt}, we use a layer of MLP and the softmax function as the classification model. The cross-entropy loss is applied to train the classification model. In real application scenarios, the labeled data are usually scarce. Following~\cite{hu2020gpt}, on each dataset, we randomly choose 10\% of items for training (\ie fine-tuning) the classification model, and the remaining 90\% of items are used for testing. In the testing phase, the model predicts the probability that an item belongs to each label, and then sorts the labels according to the predicted probabilities in descending order. The item classification performances are evaluated by Recall@5, NDCG@5 (respectively denoted by R@5 and N@5), and AUC. For each metric, we first compute the accuracy for each testing item and then report the averaged accuracy over all the testing items.

\subsection{Performance Comparison}
Table~\ref{tab:itemclassification} summarizes the item classification results on different datasets. We can make the following observations. Compared with baseline methods, PMGT consistently achieves the best results on all datasets in terms of all metrics. This again demonstrates that PMGT can achieve performance improvements in various downstream tasks, \eg item recommendation, CTR prediction, and item classification. Compared with PMGT-NP that replaces the two pre-training tasks with the classification task, the pre-trained PMGT model achieves much better performances on all datasets. This observation indicates that a pre-trained model can capture more information to help downstream tasks.

\section{Case Study in Online Platform}
\label{sec:case}

A case study is conducted in the video recommendation scenario of Taobao, one of the largest E-commerce platforms in China. The video graph is built based on users' watching behaviors. Let $r_{ht}$ denote the number of users who have watched the videos $h$ and $t$ within one hour. If $r_{ht} \geq 10$, we build an edge between the nodes $h$ and $t$ in the video graph. The weight of the edge $e_{ht}$ between $h$ and $t$ is defined following Eq.~\eqref{eq:edgeweight}. Finally, there are about 4 million nodes and 500 millions of edges in the video graph used for this case study. Given the pre-trained video representations by PMGT, for a user, we retrieve 50 most similar videos for each video she has watched, based on the Cosine similarity between the video representations. Then, ItemKNN is used to rank and recommend the retrieved videos to the user. After three days of online testing, for 600 thousand users, the number of new video plays increases by 6.80\%, compared with the online baseline method using the video representation learned by UNITER~\cite{chen2019uniter}. 
Figure~\ref{fig:case_study} shows two video retrieval examples. We can note that the videos retrieved based on the representations pre-trained by PMGT are more diverse than those retrieved based on the representations pre-trained by UNITER. 

%

\section{Conclusion and Future Work}
\label{sec:conclusion}
This paper proposes a novel pre-training GNN framework, named PMGT (\ie \underline{P}re-trained \underline{M}ultimodal \underline{G}raph \underline{T}ransformer), which exploits items' multimodal information guided by the unsupervised learning tasks on graph. Two graph reconstruction tasks, \ie graph structure reconstruction and masked node feature reconstruction, are used as learning objectives to pre-train the model. The learned representation of an item not only integrates the multimodal information of the item itself but also aggregates the information of its contextual neighbors in the graph. The superiority of PMGT has been validated by three downstream tasks (\ie item recommendation, CTR prediction, and item classification) on real datasets. In this work, we focus on the homogeneous graph of items. For future work, we would like to investigate how to extend the proposed model to process the heterogeneous item graph.

\bibliographystyle{IEEEtran}

\bibliography{documents}

\end{document}